\documentstyle[epsfig,aps,prl]{revtex}
\newcommand{\kl}{\mbox{$K_{L}$}}

\newcommand{\klpiee}{\mbox{$K_{L} \rightarrow \pi^{0} e^{+} e^{-}$}}
\newcommand{\klpimumu}{\mbox{$K_{L} \rightarrow \pi^{0} \mu^{+} \mu^{-}$}}
\newcommand{\klpinunu}{\mbox{$K_{L} \rightarrow \pi^{0} \nu \bar{\nu}$}}
\newcommand{\kleeg}{\mbox{$K_{L} \rightarrow e^{+} e^{-} \gamma $}}

\newcommand{\piz}{\mbox{$\pi^{0}$}}
\newcommand{\pizd}{\mbox{$\pi^{0}_{D}$}}

\newcommand{\pitogg}{\mbox{$\pi^{0}\rightarrow \gamma\gamma$}}
\newcommand{\pitoeeg}{\mbox{$\pi^{0}\rightarrow e^{+} e^{-}\gamma$}}

\newcommand{\meeg}{\mbox{$m_{ee\gamma}$}}

\newcommand{\kltwopzd}{\mbox{$K_{L} \rightarrow \pi^{0}\pi^{0}_{D}$ }}
\newcommand{\klthreepzd}{\mbox{$K_{L}\rightarrow\pi^{0}\pi^{0}\pi^{0}_{D}$ }}
\newcommand{\klpmzd}{\mbox{$K_{L}\rightarrow\pi^{+}\pi^{-}\pi^{0}_{D}$ }}

\newcommand{\kethree}{\mbox{$K_{e3}$}}

\newcommand{\lamnpzd}{\mbox{$\Lambda^{0} \rightarrow n \pi^{0}_{D}$ }}
\newcommand{\caspznpzd}{\mbox{$\Xi^{0} \rightarrow \Lambda^{0} (\rightarrow n \pi^{0}_{D}) \pi^{0}$ }}
\newcommand{\caspzdppi}{\mbox{$\Xi^{0} \rightarrow \Lambda^{0} (\rightarrow p \pi^{-}) \pi^{0}_{D}$ }}
\newcommand{\caspzdnpz}{\mbox{$\Xi^{0} \rightarrow \Lambda^{0} (\rightarrow n \pi^{0}) \pi^{0}_{D}$ }}

\newcommand{\caslampzd}{\mbox{$\Xi^{0} \rightarrow \Lambda^{0} \pi^{0}_{D}$}}
\newcommand{\itap}{\mbox{$p$} }
\newcommand{\itapt}{\mbox{$p_{t}$}}

\begin{document}
%
%

\title{
Search for the Decay
$K_{L} \rightarrow \pi^{0} \nu \overline{\nu}$
using $\piz \rightarrow e^{+} e^{-} \gamma$
}

\draft

\author{
A.~Alavi-Harati$^{12}$,
I.F.~Albuquerque$^{10}$,
T.~Alexopoulos$^{12}$,
M.~Arenton$^{11}$,
K.~Arisaka$^2$,
S.~Averitte$^{10}$,
A.R.~Barker$^5$,
L.~Bellantoni$^7$,
A.~Bellavance$^9$,
J.~Belz$^{10}$,
R.~Ben-David$^7$,
D.R.~Bergman$^{10}$,
E.~Blucher$^4$, 
G.J.~Bock$^7$,
C.~Bown$^4$, 
S.~Bright$^4$,
E.~Cheu$^1$,
S.~Childress$^7$,
R.~Coleman$^7$,
M.D.~Corcoran$^9$,
G.~Corti$^{11}$, 
B.~Cox$^{11}$,
M.B.~Crisler$^7$,
A.R.~Erwin$^{12}$,
R.~Ford$^7$,
A.~Glazov$^4$,
A.~Golossanov$^{11}$,
G.~Graham$^4$, 
J.~Graham$^4$,
K.~Hagan$^{11}$,
E.~Halkiadakis$^{10}$,
K.~Hanagaki$^{8,\dagger}$,
M.~Hazumi$^8$,
S.~Hidaka$^8$,
Y.B.~Hsiung$^7$,
V.~Jejer$^{11}$,
J.~Jennings$^2$,
D.A.~Jensen$^7$,
R.~Kessler$^4$,
H.G.E.~Kobrak$^{3}$,
J.~LaDue$^5$,
A.~Lath$^{10}$,
A.~Ledovskoy$^{11}$,
P.L.~McBride$^7$,
A.P.~McManus$^{11}$,
P.~Mikelsons$^5$,
E.~Monnier$^{4,*}$,
T.~Nakaya$^7$,
U.~Nauenberg$^5$,
K.S.~Nelson$^{11}$,
H.~Nguyen$^7$,
V.~O'Dell$^7$, 
M.~Pang$^7$, 
R.~Pordes$^7$,
V.~Prasad$^4$, 
C.~Qiao$^4$, 
B.~Quinn$^4$,
E.J.~Ramberg$^7$, 
R.E.~Ray$^7$,
A.~Roodman$^4$, 
M.~Sadamoto$^8$, 
S.~Schnetzer$^{10}$,
K.~Senyo$^8$, 
P.~Shanahan$^7$,
P.S.~Shawhan$^4$, 
W.~Slater$^2$,
N.~Solomey$^4$,
S.V.~Somalwar$^{10}$, 
R.L.~Stone$^{10}$, 
I.~Suzuki$^8$,
E.C.~Swallow$^{4,6}$,
R.A.~Swanson$^{3}$,
S.A.~Taegar$^1$,
R.J.~Tesarek$^{10}$, 
G.B.~Thomson$^{10}$,
P.A.~Toale$^5$,
A.~Tripathi$^2$,
R.~Tschirhart$^7$, 
Y.W.~Wah$^4$,
J.~Wang$^1$,
H.B.~White$^7$, 
J.~Whitmore$^7$,
B.~Winstein$^4$, 
R.~Winston$^4$, 
J.-Y.~Wu$^5$,
T.~Yamanaka$^8$,
E.D.~Zimmerman$^4$
}


\address{
$^1$ University of Arizona, Tucson, Arizona 85721 \\
$^2$ University of California at Los Angeles, Los Angeles, California 90095 \\
$^{3}$ University of California at San Diego, La Jolla, California 92093 \\
$^4$ The Enrico Fermi Institute, The University of Chicago, 
Chicago, Illinois 60637 \\
$^5$ University of Colorado, Boulder, Colorado 80309 \\
$^6$ Elmhurst College, Elmhurst, Illinois 60126 \\
$^7$ Fermi National Accelerator Laboratory, Batavia, Illinois 60510 \\
$^8$ Osaka University, Toyonaka, Osaka 560 Japan \\
$^9$ Rice University, Houston, Texas 77005 \\
$^{10}$ Rutgers University, Piscataway, New Jersey 08855 \\
$^{11}$ The Department of Physics and Institute of Nuclear and 
Particle Physics, University of Virginia, 
Charlottesville, Virginia 22901 \\
$^{12}$ University of Wisconsin, Madison, Wisconsin 53706 \\
$^{*}$ On leave from C.P.P. Marseille/C.N.R.S., France \\
$^{\dagger}$ To whom correspondence should be addressed. kazu@fnal.gov \\
}

\maketitle
%
%
%
\begin{abstract}

We report on a search for the decay \klpinunu, carried out as a
part of E799-II, a rare \kl\ decay experiment at Fermilab.
Within the Standard Model, the \klpinunu\ decay is dominated by
direct CP violating processes, and thus an observation of the decay
implies confirmation of direct CP violation.
Due to theoretically clean calculations, a measurement of
$B(\klpinunu)$ is one of the best ways to determine the CKM parameter 
$\eta$.
No events were observed, and we set an upper limit
$B(\klpinunu) < 5.9 \times 10^{-7} $
at the 90\% confidence level.

\end{abstract}

\newpage
\narrowtext
\twocolumn
%
%
%
The decay \klpinunu\ is dominated by direct CP violating processes
within the Standard Model through second order diagrams such as Z
penguins \cite{ref:Littenberg}.
Indirect CP violating and CP conserving contributions are expected to
be highly suppressed
\cite{ref:Littenberg2,ref:CPconserve,ref:Ritchie,ref:Hanagaki}
for the following reasons.
First order decay diagrams, which lead to relatively large
indirect CP violation in $\kl \rightarrow \pi\pi$, do not contribute
to $K_{L,S} \rightarrow \piz \nu \bar{\nu}$
because of the absence of tree level flavor changing neutral current.
Indirect CP violating contribution via second order diagrams is
suppressed by five order of magnitude ($\epsilon^{2}$).
Long-distance indirect CP violating and CP conserving contributions
from $\kl \rightarrow \piz \gamma^{*}$ and 
$\kl \rightarrow \piz \gamma^{*} \gamma^{*}$ intermediate states,
which are significant in \klpiee\ and \klpimumu, do not exist in
\klpinunu\ because the neutrinos in the
final state do not couple to virtual photons \cite{ref:Sehgal}.

Following the Wolfenstein parametrization of the CKM matrix
\cite{ref:CKM,ref:Wolfenstein}, $B(\klpinunu)$ is
proportional to $\eta^{2}$.
The uncertainty of the hadronic matrix element in
\klpinunu\ is eliminated by the experimental measurement of
$\Gamma(K^{+} \rightarrow \piz e^{+} \nu)$ and the lifetime of \kl,
which leads to an 
uncertainty of $\pm 1.5$\% in the expectation of $B(\klpinunu)$.
In addition, due to the small uncertainty ($\sim 3$\%) in the
next-to-leading order QCD correction \cite{ref:Buras},
$B(\klpinunu)$ gives direct access to $\eta$.
The current knowledge of the CKM parameters\cite{ref:CKM_limit}
allows us to predict $B(\klpinunu)$ to be $(1 \sim 5) \times
10^{-11} $\cite{ref:BR_klpinunu}.
The uncertainty comes directly from the input CKM parameters.
As the theoretical calculations are unambiguous, an observation of the
decay \klpinunu\ at the sensitivity of $\sim 10^{-11}$ would indicate
the existence of direct CP violation, and an observation outside the
predicted range would indicate new physics \cite{ref:newphys}.

It is experimentally difficult to search for \klpinunu\ because the
signature is only an isolated \piz.
The current upper limit, 
$B(\klpinunu) < 1.6 \times 10^{-6}$
at the 90\% confidence level, was obtained by
using \pitogg\ decay \cite{ref:TN}.
We report on the search for \klpinunu\ in the Dalitz decay
mode (\pitoeeg, \pizd) with the E799-II experiment using the KTeV
detector at Fermilab.
The data were collected in 44 days of running in 1997.
Using the Dalitz decay of \piz's made it possible to reconstruct
decay vertex position of the \piz's, allowing us to 
measure \itapt, the \piz's momentum transverse to the \kl\ beam
direction.
The \itapt\ played an important role in background suppression.

\begin{figure}[htbp]
  \begin{center}
    \leavevmode
    \epsfig{file=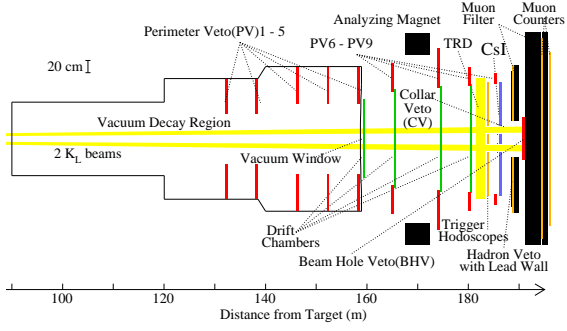,width=3.0in}
    \caption{
             The KTeV detector configuration for E799-II.
             }
    \label{fig:detector}
  \end{center}
\end{figure}

Figure~\ref{fig:detector} shows a plan view of the KTeV detector.
The elements of the detector relevant to this search are described
below.
Kaons were produced by 800~GeV proton beam, with a typical intensity
of $(3.5\sim5.0)\times 10^{12}$ protons per 19~sec beam pulse, that
struck a 30~cm 
long beryllium-oxide target at a targeting angle of 4.8~mrad.
In the first (second) half of the run period, two neutral side-by-side
beams with a solid angle of 0.25~$\mu$sr (0.35~$\mu$sr) 
each were defined by collimators downstream of the target.
A 7.6~cm long lead absorber was placed to reduce photons in the
beams.
A series of sweeping magnets removed charged particles in the beams.
The two beams entered a 69~m long evacuated decay volume starting
90~m from the target.
The vacuum was kept at $10^{-5} \sim 10^{-6}$ torr.
The downstream end of the volume was sealed by a vacuum window made of
Kevlar and Mylar.
The thickness of the vacuum window assembly was 0.0035 radiation
lengths ($X_{0}$) in total \cite{ref:vacwin}.
The neutral beam was mainly composed of neutrons and \kl's with other
long lived neutral particles, such as $\Lambda^{0}$'s and $\Xi^{0}$'s.
The relative ratios of neutron, $\Lambda^{0}$, and $\Xi^{0}$ to \kl\
at the beginning of the vacuum decay region were measured to be 3.5,
0.02 and $7.5\times 10^{-4}$, respectively.
The average kaon momentum was 70~GeV/c.
Approximately 3\% of the kaons decayed inside the vacuum decay
region.

The position and momentum of charged particles were measured using a
spectrometer consisting of four drift chambers, two upstream and two
downstream of a dipole analyzing magnet.
The magnet had a momentum kick of 205~MeV/c.
Each chamber consisted of two orthogonal views ($x$ and $y$), and had
approximately 100~$\mu$m single-hit position resolution per view.
An electromagnetic calorimeter with dimensions of 1.9~m $\times$ 1.9~m
and 27~$X_{0}$ in depth was used for photon detection and electron
identification \cite{ref:CsI}.
It was composed of 3100 pure CsI crystals.
The calorimeter had two $\rm 15 \times 15~cm^{2}$ holes located near
the center of the array to allow neutral beams to pass through.
The energy resolution of the calorimeter was below 1\%
averaged over the electron energy range 2 to 60~GeV.
A scintillator hodoscope was placed just upstream of the calorimeter
for charged particle triggering.
There were 8 Transition Radiation Detectors (TRD's) 
between the spectrometer and the trigger hodoscope for e/$\pi$
separation.
The TRD's consisted of polypropylene fiber mats as radiators and
active MWPC volumes.

The hermetic photon veto system detected photons missing the fiducial
area of the calorimeter.
The system consisted of 3 sets of counters:
perimeter vetoes (PV) 1-9, a collar veto (CV) and a beam hole
veto (BHV).
Each photon veto counter had a sandwich structure of lead (tungsten in
the CV) and scintillator.
The total depth of radiator was 16~$X_{0}$ for PV's,
8.6~$X_{0}$ for CV, and 30~$X_{0}$ (equivalent to $\sim$1 nuclear
interaction length) for BHV.
The PV covered the outer part of the calorimeter and the fiducial
volume.
The CV was placed just upstream of the calorimeter and around the two
beam holes.
The BHV was located downstream of the calorimeter and in the neutral
beam region.
The BHV was segmented into two transverse sections (one per beam) and
three longitudinal sections (10~$X_{0}$ each).
The first longitudinal section was designed to detect photons, and the
last one to detect neutrons.
Downstream of the calorimeter, there was a 10~cm lead wall followed by
a scintillator plane (hadron veto) to reject charged pions.

The trigger was designed to accept events with two electrons
and a photon so that $\kl \rightarrow \pizd\nu\bar{\nu}$
and \kleeg\ decays were accepted.
The \kleeg\ decays were used to measure the number of decayed \kl's.
The trigger hodoscope and drift chambers were used to select two
charged track events.
The calorimeter was required to have an energy deposit greater
than 18 (24)~GeV in the first (second) part of the running period.
Events with significant energy in the photon or hadron vetoes were
rejected.
Events with three or four clusters in the calorimeter with a minimum
energy of 1~GeV were selected by the hardware cluster counting
system \cite{ref:hcc}.
The TRD pulse height information was used to identify electrons at
trigger level.

The strategy in offline selection was to identify \pizd\ decays
by reconstructing the invariant mass (\meeg) and selecting high
\itapt\ events in order to suppress backgrounds.
The \itapt\ cut was used because \piz's from \klpinunu\ have a higher
kinematic \itapt\ limit than those from most of background processes.

In order to avoid human bias in the determination of selection
criteria, a blind analysis was performed.
A masked region was defined in the \itapt\ vs \meeg\ plane
as $125<\meeg(\rm MeV/c^{2})<145$ and $160<\itapt(\rm MeV/c)<240$.
Monte Carlo (MC) simulation was used to optimize all cuts while
data within the masked region were hidden.

The offline event selection began with the identification of \pizd\
decays by requiring $\rm 125 < \meeg(MeV/c^{2}) < 145$
$(\sim\pm3\sigma)$.
There were five categories in the remaining backgrounds such as:
$\kl \rightarrow \pi^{\pm} e^{\mp} \nu$ (\kethree); \klpmzd; hyperon
decays; $\kl \rightarrow \piz \pizd$ and \klthreepzd; beam
background.
Below we describe the cuts to suppress each background.

One serious background was \kethree\ decays where a photon was
radiated from the electron or overlapped accidentally, and the pion
was misidentified as an electron.
Electrons were selected by requiring $0.95<E/\itap<1.05$ where $E$ is
the energy deposited in the calorimeter and $p$ is the momentum
measured by the spectrometer.
This cut was 94\% efficient for detecting both electrons and 0.4\% for
a pion.
The transverse shower shape at the calorimeter was also used to
distinguish electrons from pions.
The confidence level to identify pions formed from the 8 TRD's
was required to be less than 1\%, which gave a 95.0\% efficiency for
electrons.
Events with out-of-time accidental energy in the calorimeter were
rejected.
The photon energy was required to be greater than 3~GeV because
accidental and radiated photons typically have lower energy.
Dalitz decays, which favor low $m_{ee}$, were selected by requiring
$m_{ee}/m_{ee\gamma} < 0.3$, 
where $m_{ee}$ is the invariant mass of the electron pair.
Defining $\theta_{+} (\theta_{-})$ as the angle between a photon and
a positron (electron) in the kaon rest frame, $\cos \theta_{+}
+ \cos \theta_{-}$ was required to be less than -1.5, because
$\pi^{\pm}$ and $e^{\mp}$ in semileptonic decays prefer a wide opening
angle, thus a peak of zero in the $\cos \theta_{+} + \cos \theta_{-}$
distribution.
These two kinematic cuts rejected 99.6\% of \kethree\ events
with a signal efficiency of 78\%.

Backgrounds involving \pizd\ decays with unreconstructed charged
particles, such as \klpmzd,  were suppressed by eliminating events
with more activity in
the drift chambers than expected from two charged track events.

High momentum $\Lambda^{0}$'s and $\Xi^{0}$'s
could reach the decay region in spite of their short life time.
Decays of these hyperons could lead to backgrounds such as \lamnpzd\
and \caslampzd, because of undetected neutrons.
These backgrounds were reduced by requiring the $z$ position, or decay
distance from the target, to be greater than 120~m.
Since hyperons decaying in the decay region had higher energy than
kaons, typically 200 to 300 GeV/c, events with photon energy greater
than 50~GeV were rejected.
In order to suppress backgrounds with neutrons such as \lamnpzd, the
energy deposited to the third segment of BHV was required to be less
than 200 minimum ionizing particles equivalent.
This cut was applied only for the $+x (-x)$ side of BHV when the decay
vertex was found in the $+x (-x)$ region to minimize the signal loss
due to accidental activity.

The $\kl \rightarrow \piz \pizd$ and \klthreepzd\ backgrounds were
suppressed by the photon veto system.
The thresholds for measured photon energy were
set to 200~MeV for PV1 and PV2, 250~MeV for PV3,
100~MeV for the rest of PV's, 1~GeV for CV, and
5~GeV (8.5~GeV) for the first section of BHV on the same (opposite)
side as the reconstructed decay position.
The electromagnetic calorimeter was also used as a part of the photon
veto system.
The number of clusters with energy greater than 1~GeV was
required to be three, and events with extra clusters with
energy greater than 250~MeV were rejected.
The photon veto requirements rejected 99.5\% of \kltwopzd\
and over 99.99\% of \klthreepzd\ events, while 
41\% of the signals, as measured by \kleeg\ events, was lost.
Most of the signal loss arose from the BHV, because the high rate
neutral beams (13~MHz \kl\ and 44~MHz neutron) struck the BHV.

\begin{figure}[htbp]
  \begin{center}
    \leavevmode
    \epsfig{file=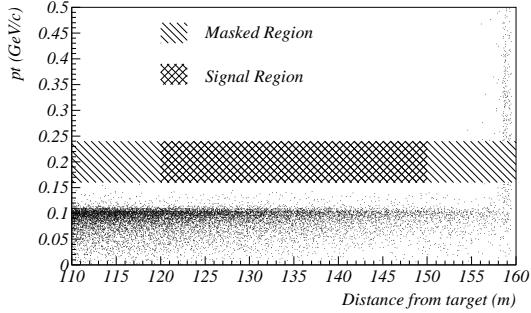,width=3.0in}
    \caption{
             The \itapt\ vs $z$ before the \itapt\ cut.
             The location of the vacuum window is $z$ = 159~m.
             }
    \label{fig:ptzvt}
  \end{center}
\end{figure}

Another background was associated with \piz's produced by beam
interactions with detector materials, primarily the vacuum window.
Figure~\ref{fig:ptzvt} shows the \itapt\ vs $z$ distribution for
data.
There was a cluster of events at $z\simeq 159$~m, the
location of the vacuum window.
To reject such events, the decay vertex position in $z$ was required
to be less than 150~m.

\begin{figure}[htbp]
  \begin{center}
    \leavevmode
    \epsfig{file=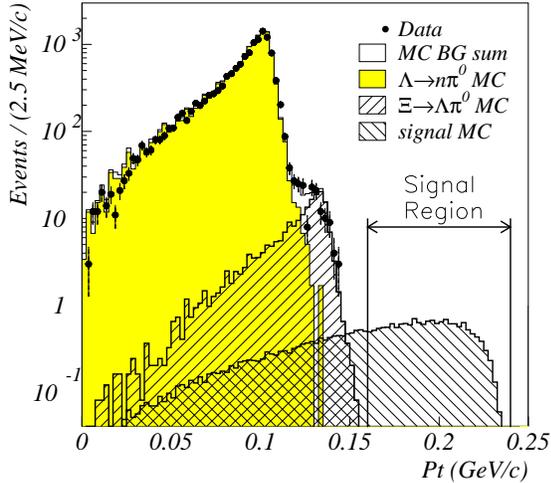,width=2.9in}
    \caption{
             The final \itapt\ distribution.
             The dots represent data, and open histogram is for MC
             expectation.
             Two main background contributions are overlaid.
             Also shown is the signal distribution predicted from the
             MC whose normalization is arbitrary.
             }
    \label{fig:pt}
  \end{center}
\end{figure}

The remaining backgrounds were primarily from hyperon decays, which
had a well-reconstructed \pizd\ decay in the fiducial region.
These were rejected by requiring \itapt\
to be $\rm 160 < \itapt(MeV/c) < 240$ as shown in Figure~\ref{fig:pt}.
The cut on the high end was determined from the kinematic limit of
\klpinunu\ decays, allowing for resolution.
The main peak arose from \lamnpzd, and the shoulder at 135~MeV/c was
from \caslampzd.
The MC events were normalized by the absolute
number of decayed \kl's, $\Lambda^{0}$'s and $\Xi^{0}$'s.
With this absolute normalization, the agreement between data and
MC distributions is excellent.
Combining \meeg\ and \itapt\ cuts, the efficiency for \lamnpzd\ was
less than $1.4\times 10^{-6}$, and $O(10^{-2})$ to $O(10^{-5})$ for
the other $\Xi^{0}$ decay backgrounds, while the signal efficiency was
46\%.

\begin{figure}[htbp]
  \begin{center}
    \leavevmode
    \epsfig{file=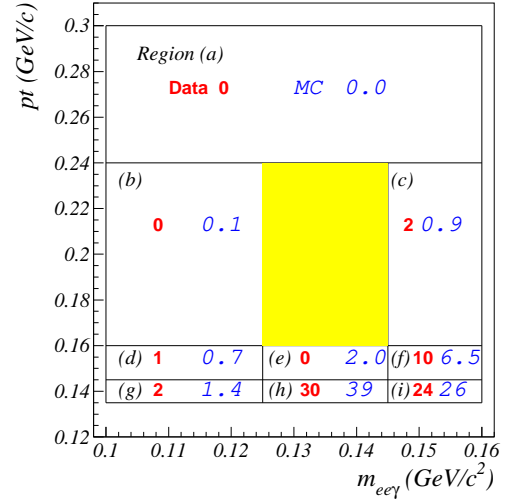,width=2.8in}
    \caption{
             The number of events around masked region.
             The left(bold) numbers represent data and the
             right(italic) is MC expectation.
             }
    \label{fig:bgshape}
  \end{center}
\end{figure}

In order to verify the MC simulation and our understanding of
the backgrounds, events around the masked region
were compared between data and MC as shown in
Figure~\ref{fig:bgshape}.
The region (f), which had the largest discrepancy of all the regions,
had a Poisson probability of 5.6\% for observing 10 events when 6.5
events were expected.
The good agreement between the expectation and the data in both the
\itapt\ shape and the number of events validates the Monte Carlo
simulation and our understanding of the backgrounds.
Even if one or more of the cuts is relaxed, the agreement is still
excellent.

Using the MC, the background levels except for those from beam
interactions were estimated and summarized in
Table~\ref{table:bglvl}. 
In the case of backgrounds associated with beam interactions, the
sideband data were used for the estimation.
Using the shape of the tail in the $z$ distribution for the 
$\itapt >$ 240~MeV/c region, as shown in Figure~\ref{fig:ptzvt}, the
contamination of backgrounds to the signal region was expected to be
0.04 events (no correlation between \itapt\ and $z$ was found).
In total, $0.12^{+0.05}_{-0.04}$ background events were expected.

The signal acceptance for \kl's decaying between 90~m
and 160~m from the target and with a momentum range of 20 to
220~GeV/c was calculated from MC to be 0.152\%.
The acceptance for \kleeg\ was similarly calculated to be 0.815\%.
The single event sensitivity (SES) of this search can be expressed as:
\begin{eqnarray*}
SES & = & \frac{1}{ A(\klpinunu) } \times \\
    &   & \frac{ A(\kleeg) }{ N(\kleeg) }
        \times \frac{B(\kleeg)}{B(\pitoeeg)} \;,
\end{eqnarray*}
where $A$, $N$, and $B$ represent the acceptance, the number of
observed events and the branching ratio of each mode, respectively.
Based on the above equation with an observed number of \kleeg\ events
of 15951 and a branching ratio of $9.1\times 10^{-6}$
\cite{ref:PDG}, the SES was calculated to be
$[2.56 \pm 0.02(stat.) \pm 0.17(sys.)] \times 10^{-7}$,
where the sources and sizes of the errors are summarized in
Table~\ref{table:syserr}.
The uncertainties on branching ratios of \kleeg\ and \pizd\
contributed a large part of the systematic error.

Finally, we examined the signal region and found no events.
Since no signal events were observed, the 90\% confidence level upper
limit on the branching ratio was determined to be
$B(\klpinunu) < 5.9 \times 10^{-7}$.

We define ``background limit'' as SES multiplied by the number of
expected backgrounds.
This figure of merit shows the experimental potential for rare decay
searches because 
it takes account not only a SES but also an expected background level.
The ``background limit'' in this search is $3.1 \times 10^{-8}$.
This is a factor of 49 lower than the ``background limit'' of
reference \cite{ref:TN}, in which the SES is $4.04\times 10^{-7}$ with
an expectation of 3.7 background events.
Even with this better technique, direct CP violation has not been
observed in this mode.
To observe the \klpinunu\ at the Standard Model level,
a search with four orders of magnitude better ``background limit''
will be required.

We gratefully acknowledge the support and effort of the Fermilab
staff and the technical staffs of the participating institutions for
their vital contributions.  This work was supported in part by the U.S. 
Department of Energy, The National Science Foundation and The Ministry of
Education and Science of Japan. 
In addition, A.R.B., E.B. and S.V.S. 
acknowledge support from the NYI program of the NSF; A.R.B. and E.B. from 
the Alfred P. Sloan Foundation; E.B. from the OJI program of the DOE; 
K.H., T.N. and M.S. from the Japan Society for the Promotion of
Science.



%
%
%
%
%
%
%
%
%
%
%
%
\begin{table}[htbp]
  \caption{
           Summary of expected background contribution in the final
           signal region.
           }
 \begin{center}
  \begin{tabular}{l|l}
  Decay mode             & Expected number of events\\
  \hline
  $K_{L}\rightarrow\pi e \nu + \gamma$          &$0.02\pm 0.02$\\
  $K_{L}\rightarrow\pi^{+}\pi^{-}\pi^{0}_{D}$   &$<0.01$\\
  $\Lambda\rightarrow n \pi^{0}_{D}$            &$<0.04$\\
  \caspzdppi                            &$0.01^{+0.006}_{-0.004}$\\
  \caspzdnpz                            &$0.01^{+0.006}_{-0.004}$\\
  \caspznpzd                                    &$0.01\pm 0.01$\\
  $K_{L}\rightarrow\pi^{0}\pi^{0}\pi^{0}_{D}$   &$0.03\pm 0.03$ \\
  $K_{L}\rightarrow\pi^{0}\pi^{0}_{D}$          &$<0.01$\\
  $n+X\rightarrow \pi^{0}X'$            &$0.04^{+0.04}_{-0.01}$\\
  \hline
  \multicolumn{1}{c|}{Total}          &$0.12^{+0.05}_{-0.04}$\\
  \end{tabular}
  \label{table:bglvl}
 \end{center}
\end{table}
%
%
\begin{table}[htb]
  \begin{center}
  \caption{
           Summary of the systematic error on the SES.
           We refer to the error coming from the statistics of \kleeg\
           events as ``statistical error'', and the remaining
           errors as ``systematic error''.
           }
  \begin{tabular}{c|c}
  Source of Error               &Size(\%)       \\
  \hline
  Statistical error             &0.79  \\
  \hline
  $B(\kleeg)$                  &5.5    \\
  $B(\pitoeeg)$                &2.7    \\
  MC statistics                 &0.81  \\
  Drift Chamber inefficiency    &1.89   \\
  Energy measurement            &0.80  \\
  Momentum measurement          &0.07  \\
  TRD efficiency                &1.88   \\
  \hline
  Total of systematic error     &6.78   \\
  \end{tabular}
  \label{table:syserr}
 \end{center}
\end{table}

\begin{references}

\bibitem{ref:Littenberg} L.S. Littenberg,
\em Phys. Rev. \bf D39, \rm 3322(1989).

\bibitem{ref:Littenberg2} J.S. Hagelin and L.S. Littenberg,\\
MIU-THP-89/039(1989).

\bibitem{ref:CPconserve} G. Buchalla and G. Isidori.
hep-ph/9806501(1998).

\bibitem{ref:Ritchie} J.L. Ritchie and S.G. Wojcicki,
\em Rev. Mod. Phys. \bf 65, \rm 1149(1993).

\bibitem{ref:Hanagaki} K. Hanagaki,
PhD thesis, Osaka University, 1998.

\bibitem{ref:Sehgal} D. Rein and L.M. Sehgal,
\em Phys. Rev. \bf D39, \rm 3325(1989).

\bibitem{ref:CKM} M. Kobayashi and T. Maskawa,
\em Prog. Theory. Phys. \bf 49, \rm 652(1973);
N. Cabibbo, \em Phys. Rev. Lett. \bf 10, \rm 531(1963).

\bibitem{ref:Wolfenstein} L. Wolfenstein,
\em Phys. Rev. Lett. \bf 51, \rm 1945(1983).

\bibitem{ref:Buras}
G. Buchalla and A.J. Buras,
\em Nucl. Phys. \bf B398, \rm 285(1993);
G. Buchalla and A.J. Buras,
\em Nucl. Phys. \bf B400, \rm 225(1993);
A.J. Buras,
\em Phys. Lett. \bf B333, \rm 476(1994).

\bibitem{ref:CKM_limit} P. Paganini, F. Parodi, P. Roudeau, and
A. Stocchi.
hep-ph/980229(1998);
A.J. Buras, hep-ph/9711217(1997).

\bibitem{ref:BR_klpinunu}
C. Dib, I. Dunietz and F.J. Gilman,
\em Mod, Phys. Lett. \bf A6, \rm 3573(1991);
A.J. Buras,
\em Phys. Lett. \bf B333, \rm 476(1994);
W.J. Marciano and Z. Parsa,
\em Phys. Rev. \bf D53, \rm R1(1996);
G. Buchalla and A.J. Buras,
\em Phys. Rev. \bf D54, \rm 6782(1996).

\bibitem{ref:newphys}
Y. Grossman and Y. Nir,
\em Phys. Lett. \bf B398, \rm 163(1997);
T. Hattori, T. Hasuike, and S. Wakaizumi,
hep-ph/9804412(1998).

\bibitem{ref:TN}
J. Adams et al.,
\em Phys. Lett. \bf B447, \rm 240(1999).

\bibitem{ref:vacwin}
E.D. Zimmerman,
\em Nucl. Inst. and Meth. \bf A426, \rm 229(1999).

\bibitem{ref:CsI}
A. Roodman, in 
\em Proceedings of the Seventh International Conference on Calorimetry
in High Energy Physics, \rm edited by E. Cheu et al. (World
Scientific, Singapore, 1998), \rm p. 89.

\bibitem{ref:hcc}\
C. Bown et al.,
\em Nucl. Inst. and Meth. \bf A369, \rm 248(1996).

\bibitem{ref:PDG} Particle Data Group,
\em European Phyical Journal \bf C3, \rm 1(1998).

\end{references}
\end{document}